\documentclass[12pt]{iopart}
\usepackage{graphicx}
\begin{document}
\title[Non-advective rate of advance of curved step]{Non-advective rate of
advance of curved step on smooth crystal face under steady-state conditions}
\author{Rasmus Persson\footnote{\textit{Present address}: Institute of Bioinformatics \&
Systems Biology and Department of Biological Science \& Technology, National
Chiao Tung University, Hsinchu City, Taiwan 30068, Republic of China}}
\address{Department of Chemistry \& Molecular Biology, University of
Gothenburg, Sweden, European Union}
\ead{raxp@nctu.edu.tw}

\begin{abstract}
For low to moderate supersaturations, crystals grow by lateral build-up of new
layers.  The edges of the layers are known as ``steps''. We consider the rate
of step advance on a flat crystal face under the influence of bulk diffusion in
the complete absence of advection, assuming a steady-state. In such
circumstances, the step velocity tends asymptotically to zero as the radius of
curvature increases. This counters the Gibbs-Thomson effect according to which
the rate of step advance should asymptotically increase \textit{ceteris
paribus} with increasing radius of curvature. Because of these competing
effects, the rate of step advance is expected to be non-monotonous in the
radius of curvature.
\end{abstract}


\section{Introduction}
In the understanding of crystal growth that emerged during the 20th century---in
the wake of the works by Kossel \cite{kossel27} and Stranski
\cite{stranski28}---crystals grow by the sequential attachment of their
molecular constituents
(``growth units'') to the crystal faces in a layer-by-layer build up. The point
of attachment is called the half-crystal position or, more commonly, the ``kink
site''. A molecule bound to the kink site has a binding energy half that of a
molecule in the bulk. By definition, the kink site is taken as the boundary
between the crystal and mother phases so that any molecule bound more loosely
is taken to be part of the mother phase. The crystal growth rate is thus
defined as the rate by which molecules from the mother phase are transported
and attach to kink sites. 

A major reason crystal growth is slow compared to condensation of liquids and
amorphous materials is due to the scarcity of kink sites, a fact linked to
their high energy of formation due to their high number of unsaturated bonds.
A flat crystal face with no defects possesses no kink sites and may thus, in
theory, not grow. Real crystal faces, however, are always somewhat defective
and thus growth is possible. A flat crystal face will exhibit clear molecular
terraces during low to moderate rates of growth. These are one molecule thick
and their edges are known as \emph{crystal steps}. It is on these steps where
the kink sites are to be found, and the basic problem in calculating the rate
of crystal growth is reduced to calculating the rate at which these steps
advance over the crystal face.

While it can be argued that between surface (2D) and bulk (3D) diffusion there
is no clear-cut physical distinction, the former being geometrically
\emph{contained} in the latter, the division has found great practical value in
the mathematical treatment. Burton, Cabrera and Frank \cite{burton51} solved
the surface-diffusion equations for mass transport to steps. Chernov
\cite{chernov61} was the first to solve the case of pure bulk diffusion to the
straight step, in isolation or in parallel sequence. This was later generalized
by Gilmer, Ghez and Cabrera \cite{gilmer71} to account also for a diffusion of
growth units to the surface from the bulk mother phase. A general approach
involving concurrent volume and surface diffusion up to the step was presented
by van der Eerden \cite{vandereerden82}. He studied solutions for a train of
parallel, straight steps and for an isolated curved step. Common to these
treatments is the assumption that there exists a concentration gradient of
finite extension separating the bulk concentration from that at the steps: the
so-called ``unstirred boundary layer.'' Whereas in cases of growth from a
stirred solution, this is reasonable, in cases where advection is kept to a
minimum, this assumption can only be dispensed of if the curvature of the steps
is properly taken into account. This situation is of importance because in some
cases of growth from solution, stirring may be undesirable.

In this Paper we solve the bulk diffusion problem up to an isolated curved step
for an arbitrarily thick diffusion layer. In the case of growth from a
well-stirred solution, the mass transport to the steps is not the rate-limiting
process. In this case, the thermodynamic stability of the steps themselves are
important factors in the rate of the crystal growth. By thermodynamic arguments
\cite{markov03}, it can be shown that the curved step is unstable due
essentially to the Gibbs-Thomson effect and thus prone to decay. Therefore, the
rate of growth for a small, nascent terrace will be lower than for the larger,
developed one which has a larger radius of curvature. However,  it will be
shown that, the Gibbs-Thomson effect notwithstanding, under bulk diffusion from
infinitely far away the step velocity should decrease after an early maximum
because mass transport becomes rate-limiting when advection is absent. The main
assumptions in the demonstration are
\begin{itemize}
\item The crystal surface is mathematically flat.
\item The step is a segment of a semitorus of greater radius $R$ and lesser
radius $a / \pi$, where $a$ is the molecular diameter.
\item The step velocity is very slow compared to that of the incoming growth
units.
\item Activity coefficients are unity. 
\item There is a steady-state.
\item Surface diffusion has the same diffusion constant as that of bulk and
hence there is no Ehrlich-Schwoebel barrier \cite{ehrlich66,schwoebel66} (the
unequal mass transport to a step due to the inherent asymmetry in the potential
energy landscape experienced by a growth unit approaching the step over the
the terrace itself compared to approaching it from beneath). 
\item Kinks are so numerous along the step that it is essentially a continuous
sink for growth units.
\end{itemize}
Some of these assumptions may conceivably be relaxed at the expense of added
parameters but this is not expected to affect the general conclusion.

This paper is only concerned with the particular case of the rate of curved
step advance. For a more in-depth background and pedagogical account of crystal
growth and its theory, the reader is recommended to turn to the textbook by
Markov \cite{markov03}. Another excellent resource is the book by Pimpinelli
and Villain \cite{pimpinelli98}.

\section{Mass transport through steady-state diffusion}
\begin{figure}[ht]
\begin{center}
\includegraphics{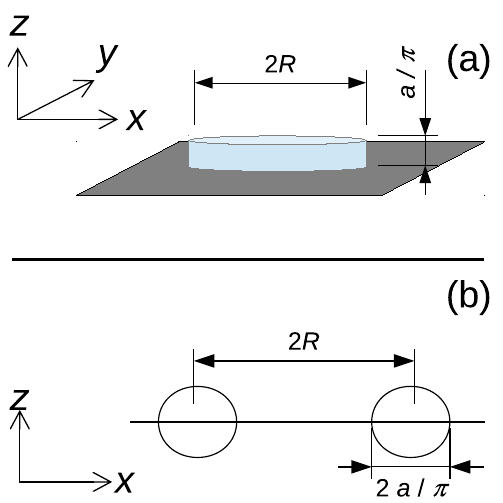}
\end{center}
\caption{Schematic of the model under consideration: A monomolecular circular
surface layer of diameter $2 R$ with a height of $a / \pi$ with the Cartesian
$z$-axis normal to the crystal plane. The origin of the Cartesian coordinate
system is at the center of the embryo. A sink keeps the concentration of growth
units within a distance $a / \pi$ from the perimeter of the embryo at a
concentration $c_1$. At infinite distance from the perimeter, the concentration
is $c_\infty$. Panel (a) shows a schematic, isometric view of the surface
embryo. Panel (b) shows a cross section of the semitorus at which the boundary
condition is applied (the region below the horizontal line is excluded from
the mass transport analysis by restriction of the toroidal $\theta$
coordinte).}
\label{fig:geom}
\end{figure}

Refer to Figure~\ref{fig:geom} for a schematic illustration of the problem. We
consider a flat crystal face coincident with the $x-y$-plane, so that the
surface normal is parallel to the $z$-axis. On the crystal face we have a
circular terrace of adsorbed growth units. The symmetry of this problem
suggests the use of toroidal coordinates. These are obtained by rotating the 2D
bipolar coordinates around a fixed axis of symmetry and may be defined as
\cite[p. 112]{moon88}
\begin{eqnarray}
x = \frac {r \sinh \eta \cos \psi} {\cosh \eta - \cos \theta} \\
y = \frac {r \sinh \eta \sin \psi} {\cosh \eta - \cos \theta} \\
z = \frac {r \sin \theta} {\cosh \eta - \cos \theta} 
\end{eqnarray}
In our case the coordinate ranges are taken to be $\eta \in [0, \infty)$,
$\theta \in [-\pi, \pi]$ and $\psi \in [-\pi, \pi]$, respectively. The
parameter $r$ is a scaling factor of the coordinate system. Its value is
arbitrary but for each value of $r$, a new toroidal coordinate system is
obtained. It is related to the greater radius of the torus, $R$, by
\begin{equation}
r = R \tanh \eta_1
\end{equation}
and to the lesser radius, which we take to be given by $a/\pi$, through
\begin{equation}
r = \frac a {\pi} \sinh \eta_1 \label{sinh}
\end{equation}
where $\eta_1$ defines the surface of the torus. This surface is what we will take
as the ``dividing surface'' of the crystal step with respect to the mother
phase. Growth units outside it are considered part of the mother phase; those
inside, part of the crystal.  The angle $\theta$ refers to the latitudinal
direction, where $\theta=\pi/2$ is directed parallel to the $z$-axis.  The
azimuthal angle $\psi$ denotes rotation about the $z$-axis at a distance 
\begin{displaymath}
\frac {r \sinh \eta} {\cosh \eta - \cos \theta}
\end{displaymath}
The steady-state diffusion without advection is described by Laplace's equation
for the concentration field. This equation is separable in toroidal coordinates
and we will cover its solution over the next subsections.

\subsection{General solution of the Laplace equation}
The Laplacian in toroidal coordinates is given by \cite[p. 112]{moon88}
\begin{equation}
\nabla^2 = \frac {(\xi - \cos \theta)^3} {r^2 (\xi - 1)} \left
[\partial_{\theta} \left ( \frac {(\xi - 1) } {\xi - \cos \theta}
\partial_{\theta} \right ) + \partial_{\eta} \left (\frac {(\xi - 1)} {\xi -
\cos \theta} \partial_\eta \right ) + \frac {(\xi-1)^{-1} \partial_{\psi} ^2}
{\xi - \cos \theta} \right ]
\end{equation}
where $\xi = \cosh \eta$. Denoting the concentration of growth units by $c(\xi,
\theta, \psi)$, the Laplace equation is
\begin{equation}
\label{eq:laplace}
\nabla^2 c(\xi, \theta, \psi) = 0 
\end{equation}
This equation is separable and if $c(\xi, \theta, \psi)$ is expressed
as \cite[pp. 112-115]{moon88}
\begin{equation}
c(\xi, \theta, \psi) = A + \sqrt{\xi - \cos \theta} H(\eta) \Theta(\theta)
\Psi(\psi)
\end{equation}
where $A$ is a constant, the solutions are obtained from
\begin{eqnarray}
(\xi^2 - 1)H'' + 2 \xi H' - ((p^2 - 1/4) + q^2/(\xi^2 - 1)) H  =  0 \\
\Theta'' + p^2 \Theta  =  0 \label{eq:theta} \\
\Psi'' + q^2 \Psi  =  0 \label{eq:psi}
\end{eqnarray}
where the primes indicate differentiation. The first of these is a variant of
the associated Legendre equation ($p, q$ are constants) and its solutions are
linear combinations of the associated Legendre functions \cite{abramowitz64}
$\{P_{p-\frac 1 2}^{q}(\xi)\}$ and $\{Q_{p-\frac 1 2}^q(\xi)\}$. However, the
$Q$-functions diverge for $\eta \to 0$, corresponding to great distance from
the torus or close to the $z$-axis, and must be excluded for physical reasons. 

Eqs~(\ref{eq:theta}) and~(\ref{eq:psi}) are less intimidating in that they
share their mathematical form with the equation-of-motion of the undamped
harmonic oscillator. The solutions are well-known and, as in the case of the
oscillator, for physical reasons of continuity, the solutions have to be
periodic around the torus, so that the general form is constrained to
\begin{eqnarray} 
\Theta(\theta) = B_p \cos (p \theta) + C_p \sin (p \theta) \\ 
\Psi(\psi) = B_q' \cos(q \psi) + C_q' \sin (q \psi) 
\end{eqnarray}
with $p, q$ integers and $B_p, C_p, B_q', C_q'$ arbitrary constants. This set
of solutions is more general than we need, however, because we assume azimuthal
symmetry and so we may take $q = 0$ and let $\Psi(\psi) \equiv 1$ without any
other loss of generality. In this case, the general solution reduces to
\begin{equation}
c(\xi, \theta) = A + \sqrt{\xi - \cos \theta} \sum_{p=0} ^{\infty}
(B_p \cos(p \theta) + C_p \sin(p \theta)) P_{p-\frac 1 2}(\xi) 
\end{equation}
with all $\psi$-dependence suppressed.

\subsection{Particular solution for a continuous step at arbitrary distance from
bulk in radial direction}
The coordinate $\xi_1$ defines the surface of the step where the concentration
is
\begin{equation}
c(\xi_1, \theta) \equiv c_1
\end{equation}
whereas $\xi_{\delta} < \xi_1$ defines a toroidal surface at a distance
$\delta$ from the step (the unstirred boundary layer), where the concentration
has reached its bulk value,
\begin{equation}
c(\xi_{\delta}, \theta) \equiv c_{\infty} \\
\end{equation}
This means that
\begin{equation}
c_1 = A + \sqrt{\xi_1 - \cos \theta} \sum_{p=0} ^{\infty} (B_p
\cos(p \theta) + C_p \sin(p \theta))P_{p-\frac 1 2}(\xi_1)
\end{equation}
and
\begin{equation}
c_{\infty} = A + \sqrt{\xi_{\delta} - \cos \theta} \sum_{p=0} ^{\infty}
(B_p \cos(p \theta) + C_p \sin(p \theta)) P_{p-\frac 1 2}(\xi_{\delta}) \\
\end{equation}
With two boundary conditions and three unknowns ($A$, $B_p$, and
$C_p$), we need a further condition and impose the physically reasonable case
that $c_{\infty}$ is even in $\theta$, in which case all sine terms must
vanish, \textit{i. e.} $C_p = 0, \forall p$.  Then, through the relation
(Heine's equation) \cite{andrews06}
\begin{equation}
\frac {1} {\sqrt{\xi - \cos \theta}} = \frac {\sqrt{2}} {\pi}
\sum_{p=0}^{\infty} (2 - \delta_{0p}) Q_{p-\frac 1 2}(\xi) \cos(p \theta),
\end{equation}
where $\delta_{0p}$ designates the Kronecker delta, we may deduce that
\begin{equation}
B_p = \frac {\sqrt{2} (c_1 - A) (2 - \delta_{0p})} {\pi} \frac
{Q_{p-\frac 1 2}(\xi_1)} {P_{p-\frac 1 2}(\xi_1)} \label{eq:bps}
\end{equation}
and
\begin{equation}
B_p = \frac {\sqrt{2} (c_{\infty} - A) (2 - \delta_{0p})} {\pi} \frac
{Q_{p-\frac 1 2}(\xi_{\delta})} {P_{p-\frac 1 2}(\xi_{\delta})}
\end{equation}
Summing both equations and then solving for $A$, yields
\begin{equation}
\label{eq:aps}
A = \frac {c_1 \Sigma_1(\xi_1) - c_{\infty} \Sigma_1(\xi_{\delta})}
{\Sigma_1(\xi_1) - \Sigma_1(\xi_{\delta})}
\end{equation}
where we have defined
\begin{equation}
\Sigma_1(x) = \sum_{p=0}^{\infty} \frac {\sqrt{2}(2 - \delta_{0p})} {\pi}
\frac {Q_{p-\frac 1 2}(x)} {P_{p-\frac 1 2}(x)}
\end{equation}
for brevity.

Finally, we have
\begin{equation}
\label{eq:inv1}
\xi_1 = R \pi / a
\end{equation}
and, if $\delta \ll R$, \emph{i. e.} for a sufficiently great radius of
curvature,
\begin{equation}
\xi_{\delta} = R / \delta 
\end{equation}
otherwise, $\forall x, y, z | x^2 + y^2 + z^2 = \delta^2$, it is
\begin{equation}
\label{eq:lim}
\xi_{\delta} = \cosh \left ( \ln \left (\sqrt {\frac {(\sqrt{x^2 + y^2} + r)^2 + z^2}
{(\sqrt{x^2 + y^2} - r)^2 + z^2}} \right ) \right )
\end{equation}
by virtue of the inverse coordinate transformation. It is to be noted from this
last expression that $\xi_{\delta} \to 1$ when $\delta \to \infty$.

\section{Steady-state step velocity}
Having established the concentration field function, we may compute the
steady-state diffusion flux density to the step through Fick's first law.
In toroidal coordinates, the gradient operator is given by
\begin{equation}
\nabla = \frac {\xi - \cos \theta} {r} \left (\widehat{\eta} \frac
{\partial} {\partial \eta} + \widehat{\theta} \frac {\partial} {\partial
\theta} + \frac {\widehat{\psi}} {\sinh \eta} \frac {\partial} {\partial \psi}
\right ),
\end{equation}
where a circumflex indicates a unit vector parallel to the indicated axis.
Since we are interested only in the flux across the toroidal surface---mass transport
along the $\theta$ direction is not taken to be integrated into the lattice---the
diffusion flux density depends only on the $\eta$-derivative and we have
\begin{equation}
j_{\mathrm d}(\theta) = -\frac {D(\xi_1 - \cos \theta)} {r} \left ( \frac
{\partial c} {\partial \eta} \right )_{\eta=\eta_1}
\end{equation}
where $D$ is the gradient diffusion constant of the growth units \footnote{Note
that if $D$ is taken to be $\theta$-dependent, we can include into the
formalism both the Ehrlich-Schwoebel barrier and surface diffusion.}.  The
dependence on $\theta$ in this equation is unlikely to be physically
realistic, because it stems from purely geometric facts and does not account
for the Ehrlich-Schwoebel barrier. However, we need not consider the flux in
detail, but may rather think of the aggregate flux density as integrated over
all $\theta$ angles. At steady-state, the flux density reaching the step by
mass transport is precisely balanced by the integration flux density of growth
units being incorporated into the crystal lattice. Denoting the integration
flux density by $j_\mathrm i$, this steady-state condition reads
\begin{equation}
\label{eq:steadystate}
\int_{0}^{\pi} j_\mathrm{d}(\theta) \mathrm d\theta = j_\mathrm i
\end{equation}
where the integration over $\theta$ only covers the angular range corresponding
to the semitorus above the crystal plane since no mass transport from the crystal
plane itself can contribute. 

Typically (see \textit{e. g.} Ref.~\cite{markov03}), $j_\mathrm i$ is written
as the product of an integration frequency $\nu$, the concentration difference
at the step with respect to the solubility $c_1 - c_0$, and the density of kink
sites $a^2 / x_0$ ($x_0$ is the mean distance between kinks and $a^2$ the mean
cross-sectional area of a growth unit), thus
\begin{equation}
j_\mathrm i = \frac {a^2} {x_0} (c_1 - c_0) \nu
\end{equation}
The step velocity (with sufficient advection) as a function of the radius of
the terrace is to lowest non-vanishing order in $R^* < R$ given by
\cite{markov03}
\begin{equation}
\label{eq:vel}
v_{\infty}(R) = v_{\mathrm m} j_\mathrm i (1 - R^*/R)
\label{stepvel}
\end{equation}
where $v_{\mathrm m}$ is the molecular volume of the growth unit in the
crystal, and $R^*$ is a characteristic radius of stability for the surface
terrace which inhibits the growth rate through the factor $(1 - R^* / R)$. For
small $R$, the tendency to decay through loss of growth units greatly
outweighs the tendency for growth because of an increased local chemical
potential of the terrace with respect to the mother phase due to the
destabilizing Gibbs-Thomson effect. The chemical potential difference and, by
extension, the value of $R^*$ in turn depend on the local concentration. Formation
or growth of terraces of adsorbed growth units becomes thermodynamically
unfavored when the mother phase concentration approaches the equilibrium one.
In other words, $R^*$ diverges in the limit $c_1 \to c_0$. As for the local
concentration $c_1$, its value depends on the balance between $D$ and $\nu$,
approaching $c_\infty$ as $D / \nu \to 0$ and $c_0$ as $D / \nu \to \infty$.

The increase of the step velocity with the terrace radius $R$ indicated
\textit{primae faciae} by eq.~(\ref{eq:vel}) does not hold when the dependence
of $j_\mathrm i$ on the concentration field through eq.~(\ref{eq:steadystate})
is also taken into account. It is the physical consequences of this condition
that we will now examine.

\subsection{Single step velocity in the absence of advection}
For simplicty, we consider the case of the isolated step. In this case,
\begin{equation}
\label{eq:isol}
j_{\mathrm d}(\theta) = -\frac {\pi D} {a}  \left (\frac {\Sigma_2(\theta,
\xi_1)} {2 \sqrt{\xi_1 - \cos \theta}}  + \sqrt{\xi_1 - \cos \theta} \frac
{\partial \Sigma_2(\theta, \xi_1)} {\partial \xi_1} \right ) 
\end{equation}
which is apparent using the definition---introduced for brevity---of the
function
\begin{equation}
\Sigma_2(\theta, \xi_1) = \sum_{p=0}^{\infty} B_p \cos (p \theta) P_{p - \frac
1 2}(\xi_1)
\end{equation}
and the fact that
\begin{equation}
\left (\frac {\partial \xi} {\partial \eta} \right )_{\eta = \eta_1} = \sinh
\eta_1 = \frac {\pi r} {a}
\end{equation}

Arguably, the physically most interesting case arises when we consider the
limit $\delta \to \infty$, corresponding to the complete absence of advection, as
it corresponds to the extreme influence of the mass transport on the crystal
growth rate.  With the help of eqs.~(\ref{eq:bps}), (\ref{eq:aps})
and~(\ref{eq:lim}), we see that $\Sigma_2$ turns into,
\begin{equation}
\Sigma_2(\theta, \xi_1) = \frac {\sqrt{2}} {\pi} \left (c_1 - c_\infty \right )
\sum_{p=0}^\infty (2 - \delta_{0p}) \cos (p \theta) P_{p-\frac 1 2}(\xi_1)
\end{equation}
in this limit. Moreover, since eq.~(\ref{eq:vel}) implies that $R$ is large,
and thus that $\xi_1 \gg \cos \theta$. By Taylor expansion in $\cos \theta$,
we have
\begin{equation}
j_\mathrm{d}(\theta) = -\frac {\pi D} {a}  \left \{ \frac {\Sigma_2(\theta,
\xi_1)} {2 \sqrt{\xi_1}} \left (1 + \frac {\cos \theta} {2 \xi_1^{\frac 3 2}}
\right ) + \sqrt{\xi_1} \left (1 - \frac {\cos \theta} {2 \sqrt{\xi_1}} \right )
\frac {\partial \Sigma_2(\theta, \xi_1)} {\partial \xi_1} \right \}
\end{equation}
if we truncate the expansion after the linear term. Consider now the case of
eq.~(\ref{eq:steadystate}), through which we may calculate $j_\mathrm{i}$ and,
by extension, $v_\infty(R)$. By symmetry, it is clear that
\begin{equation}
\int_{0}^{\pi} \Sigma_2(\theta, \xi_1) \mathrm d \theta = 0
\end{equation}
and because of the orthogonality of $\cos \theta$ and $\cos p \theta$ for $p
\neq 1$ over $\theta \in [0, \pi]$, we find that
\begin{equation}
\int_{0}^{\pi} \Sigma_2(\theta, \xi_1) \cos \theta \mathrm d \theta = \sqrt{2}
(c_1 - c_\infty) P_{\frac 1 2}(\xi_1)
\end{equation}
with an analogous non-vanishing result for the corresponding integral over
$\partial \Sigma_2 / \partial \xi_1$. Using mathematical reference tables or software,
the series expansion of the Legendre function (and its derivative) around
infinity reveals that the first non-vanishing asymptotic term of
$j_\mathrm{d}(\theta)$ for $R \to \infty$ goes as $R^{-1}$. Therefore, we
conclude, after taking into account eq.~(\ref{eq:vel}), that
\begin{equation}
\label{eq:svel}
v_\infty(R) \propto \frac {(c_1 - c_\infty)} {R}
\end{equation} 
asymptotically for large $R$ in the absence of advection. Although we limit
ourselves to presenting in this paper the most salient result, \textit{viz.}
the asymptotic scaling of the step velocity, the exact (within the
confines of the model) constant of proportionality before the $R^{-1}$
term, as well as arbitrary higher orders, may be found without effort using
mathematical software.

\section{Conclusion}
Within the steady-state approximation, a general equation for the isolated step
velocity with bulk diffusion from arbitrary distance, taking into account also
the inhomogeneity of the resulting diffusion field around it, has been derived.
The complexity of the expression means that in its complete form it's more
suitable for a numerical treatment, but by series expansion it was shown that
the step velocity should asymptotically vanish as the inverse radius of
curvature of the step for large values of the same. This runs directly counter
to the conclusion of the well-stirred mother phase in which the Gibbs-Thomson
effect is responsible for a monotonously increasing step velocity with
increasing radius of curvature. Therefore, one expects the rate of step advance
in the absence of advection to exhibit an early maximum because of the
competing effects of the Gibbs-Thomson instability (prominent for $R$ on the
order of $R^*$) and the steady-state mass transport restriction for the curved
step (prominent for $R \gg R^*$). 

Finally, even though by the very order of things
the effect of advection on the crystal growth rate will be
more pronounced whenever the step velocity is the rate controlling
factor---such as for growth by surface nucleation at high supersaturations
(so-called polynuclear growth where several concurrent surface clusters race to
build up the growing crystal) or in the presence of crystal screw dislocations which
serve as a persistent source of steps (so-called spiral growth)---than, for
instance, at low supersaturations with growth proceeding in the mononuclear
regime and the surface nucleation process is rate-limiting; we note that it is,
however, theoretically possible that increasing the vigor of the advection
alone may shift the growth mechanism from polynuclear to mononuclear by
increasing the average value of $v_\infty(R)$ in the system so that the surface
nucleation rate becomes limiting.

\vfill

\bibliographystyle{unsrt}
\bibliography{cg.bib}

\clearpage

\end{document}